\documentclass[aps,prl,preprint,groupedaddress]{revtex4-1}
\usepackage[USenglish]{babel} 
\usepackage{amsmath}
\usepackage{amsfonts}
\usepackage{graphicx}
\usepackage{float}
\usepackage{hyperref}
\usepackage{color}

\begin{document}
\title{Thermo-osmotic flow in thin films}

\author{Andreas P. Bregulla$^{1}$, Alois W\"urger$^{2}$, Katrin G\"unther$^{3}$, Michael Mertig$^{3,4}$, Frank Cichos$^{1}$}
\email{cichos@physik.uni-leipzig.de}
\homepage[]{http://www.uni-leipzig.de/~physik/mona}
\affiliation{$^{1}$Molecular Nanophotonics Group, Institute of Experimental Physics I, University of Leipzig, 04103 Leipzig, Germany\\
$^{2}$ Laboratoire Ondes et Mati\`ere d'Aquitaine, Universit\'e de Bordeaux \& CNRS, 33405 Talence, France\\
$^{3}$BioNanotechnology and Structure Formation Group, Department of Chemistry and Food Chemistry, Chair of Physical Chemistry, Measurement and Sensor Technology, Technische Universit\"at Dresden, 01062 Dresden, Germany\\
$^{4}$ Kurt-Schwabe-Institut f\"ur Mess- und Sensortechnik e.V. Meinsberg, 04736 Waldheim, Germany }

\begin{abstract}
We report on the first micro-scale observation of the velocity field imposed by a non-uniform heat content along the solid/liquid boundary. We determine both radial and vertical velocity components of this thermo-osmotic flow field by tracking single tracer nanoparticles. The measured flow profiles are compared to an approximate analytical theory and to numerical calculations. From the measured slip velocity we deduce the thermo-osmotic coefficient for both bare glass and Pluronic F-127 covered surfaces. The value for Pluronic F-127 agrees well with Soret data for polyethylene glycol, whereas that for glass differs from literature values and indicates the complex boundary layer thermodynamics of glass-water interfaces. 
\end{abstract}
\maketitle

Osmosis is the passage of a liquid through a semipermeable membrane, towards a higher concentration of a molecular solute or salt. Osmotic processes are fundamental for life, for example in selective transport through cell membranes, and for applications such as desalination of seawater and power generation from the salinity difference with river water \cite{Log12}. In physical terms, osmosis is driven by the gain in mixing entropy and requires to impede solute diffusion. 

Thermo-osmosis relies on the same principle, albeit with a liquid of non-uniform \emph{heat} content instead of a non-uniform solute concentration; accordingly, the underlying thermodynamic force is the temperature gradient rather than a concentration gradient. Since there are no heat-selective membranes, thermo-osmosis occurs in open geometries only, where heat and liquid flow in opposite directions along a solid boundary, similarly to electro-osmosis in capillaries or nanofluidic diodes \cite {Pic13}. Water flow due to a temperature gradient was first observed by Derjaguin and Sidorenkov through porous glass from a hot to a cold reservoir \cite{Der41}.   

In the last decade thermal gradients have become a versatile means of manipulating colloidal dispersions, e.g., self-propulsion of metal-capped Janus particles \cite{Jia10,But12}, cluster formation through hydrodynamic interactions \cite{Wei08,Leo09}, force-free steering through dynamical feedback \cite{Bre14,Qia13}, sieving by size and stretching macromolecules \cite{Duh06,Mae11,Ped14}, dynamical trapping of nano-particles \cite{Bra13,Bra15}, and detection of DNA through functionalized gold particles \cite{Yu15}.  In all these examples, the motion arises from a superposition of thermo-osmosis and molecular osmosis \cite{Wue10}: The temperature gradient induces heat flow along the colloidal surface, whereas its companion fields, e.g. composition \cite{But12,Wue15} and ion concentration \cite{Esl14,Bro14,Vig10,Rei14}, drive molecular currents. Finally we note that thermo-osmotic flow could be relevant for particle motion through hot nanostructures, in addition to thermophoresis and thermoconvection\cite{Chen15,Cuche13}. 

Here we report on the first micro-scale observation of the velocity field imposed by thermo-osmosis along the solid boundary. Both radial and vertical velocity components are determined by tracking single tracer nanoparticles. The measured flow profiles are compared to an approximate analytical theory and to numerical calculations. 

\begin{figure}[ptb]
\begin{center}
\includegraphics[width=1\columnwidth]{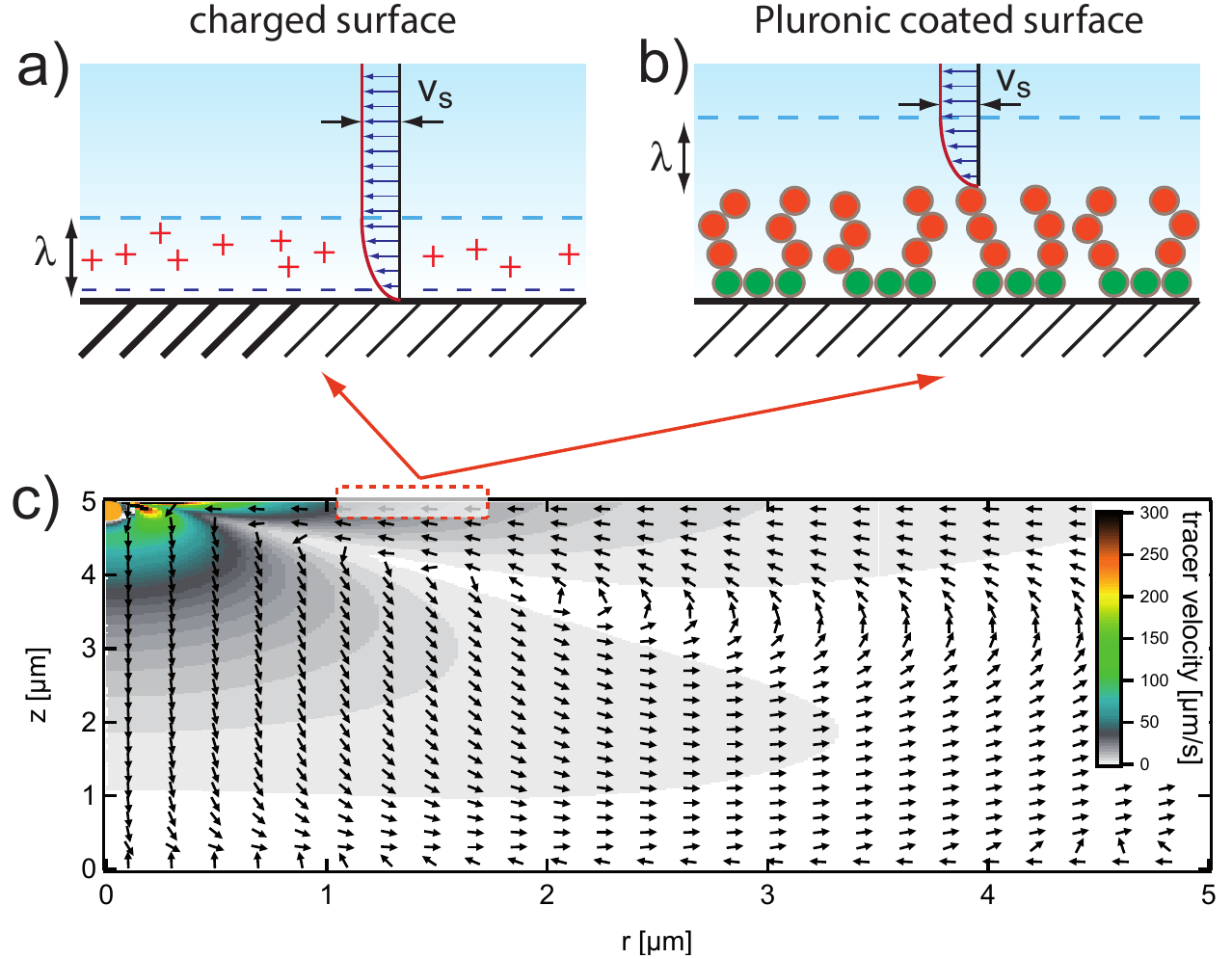}
\caption{Schematic view of the boundary layer in an interface with non-uniform
temperature. a) Thermo-osmotic velocity profile close to a charged solid boundary with an excess enthalpy within an interaction length $\lambda$; for $z>\lambda$ the velocity saturates at $v_s$. b) The same for a Pluronic-coated surface. c) Streamlines in the liquid film resulting from slip velocities at the upper and lower boundaries, reconstructed by numerical calculations based on the experimental results for the Pluronic F-127 coated surface.}
\label{fig2}
\end{center}
\end{figure}

\paragraph{Thermo-osmotic slip velocity.}
A bulk liquid in a temperature gradient reaches a non-equilibrium stationary state, with a steady non-uniform composition but zero matter flow. A solid boundary, however, exerts additional forces on the liquid; the corresponding excess (or defect) specific enthalpy $h$ in the boundary layer results in a creep flow parallel to the surface, with the effective slip velocity  \cite{Der87}
\begin{equation}
v_{s}=-\frac{1}{\eta}\int_{0}^{\infty}dz z h(z)\frac{\nabla T}{T} \equiv \chi\frac{\nabla T}{T},
\label{20}%
\end{equation}
where $\eta$ is the viscosity. An enthalpy excess in the boundary layer, $h>0$, leads to a negative $\chi$ and liquid flow towards the cold side, as observed for glass capillaries, clays, and silica gels \cite{Der41,Der87}, whereas a negative enthalpy $(h<0)$ drives the liquid towards the hot, e.g., through various synthetic membranes \cite{Kim09,Vil06}. 

In terms of Onsager's reciprocal relations, $\chi$ is the mechano-caloric cross-coefficient, which describes equally well the excess heat carried by liquid flow at constant temperature \cite{Der87}. Its explicit form can also be derived from the principles of non-equilibrium thermodynamics \cite{Gro62}: A solid boundary modifies the specific chemical potential $\mu$ within an interaction length $\lambda$. Plugging the thermodynamic force $-T\nabla(\mu/T)$ in Stokes' equation and using the Gibbs-Helmholtz relation $d(\mu/T)/dT=-h/T^2$,  one readily obtains eq. \ref{20}. Since the specific chemical potential $\mu=h-Ts$ is usually constant perpendicular to the surface, the excess enthalpy in the boundary layer is identical to the entropic term, $h=Ts$. This relates the slip velocity to the  ``entropy of transport'' $s$. 

\paragraph{Experimental} To measure the thermo-osmotic flow field we have constructed a sample cell that consists of two glass cover slides (Roth) enclosing a water film of about $5\, \rm{\mu m}$ thickness (Fig.\ \ref{fig1}a). Both slides were either used untreated or covered with Pluronic F-127 (see supplementary material for details\cite{SI}). A gold particle of radius $a=125\,\rm{nm}$ is used as a heat source to generate a well defined temperature gradient along the glass/water (polymer/water) interface (eq. \ref{eq:temperature}). The gold particle was immobilized at the upper glass surface to avoid convection.

\begin{equation}\label{eq:temperature}
	T\left(r\right)=T_{0}+\frac{P_{inc}}{4\pi\kappa r} =T_{0}+\Delta T_{\rm{Au}}\frac{a}{r} . 
\end{equation}

The gold particle is heated with a $\lambda=532\, {\rm nm}$ laser at a power of $P_{\rm inc}=5\, \rm{mW}$. 
The temperature increment of the gold particle $\Delta T_{\rm{Au}}\approx 80\,\rm{K}$ was estimated with the help of a separate experiment \cite{SI}. Fig.\ \ref{fig1} c) displays the expected temperature profiles at the liquid/solid interface corresponding to this nanoparticle temperature at the upper slide (black curve) and  at the lower slide (red curve). 
The velocity field is measured by tracking single gold nano-particles of a radius $R=75\,\rm{nm}$. Because of its high thermal conductivity, the particle is almost isothermal and thus does not migrate in the temperature gradient. As the sample cell is thicker than the focal depth ($\approx 1\, \mu m$) of the microscopy setup, the scattering intensity of the tracers varies with the $z-$position. Selecting only the brightest particles for the analysis for example allows access to the velocity field close to the solid boundary.

\begin{figure}[ptb]
\includegraphics[width=1\columnwidth]{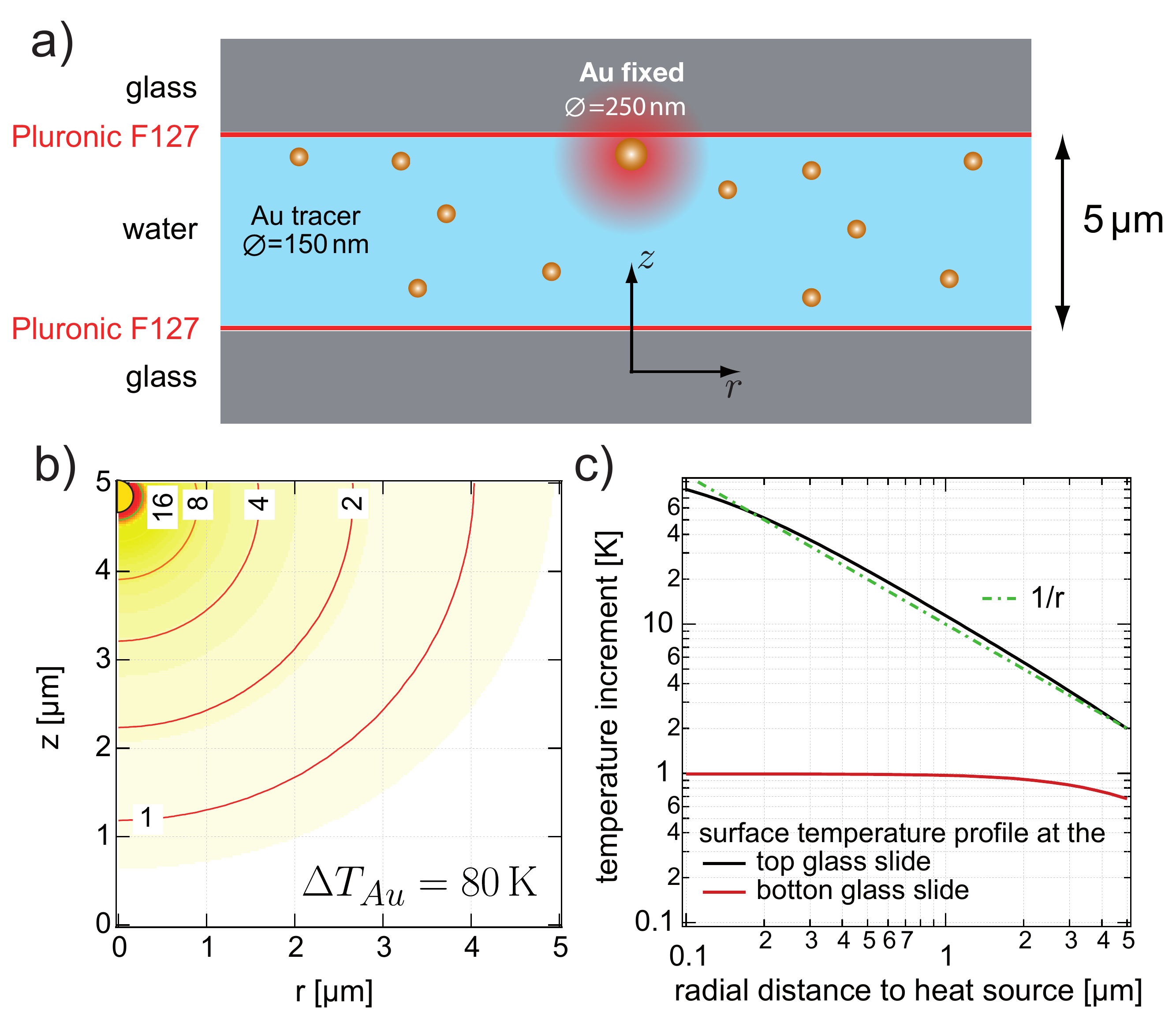}
\caption{a) Principal design of the experiment. The fluid is contained between two glass cover slides in a sandwich structure with a gap of about $5\, \rm{\mu m}$. A $R=125\, \rm{nm}$ gold particle is immobilized at the top surface and used as the heat source. A focused laser beam ($\lambda=532\,\rm{nm}$) is used to heat the particle via plasmonic heating. $R=75\,\rm{nm}$ gold tracer particles were used to measure the velocity field. b) Temperature map around the heated gold particle in cylindrical coordinates $r$ and $z$. c) Temperature profile along the upper (black curve) and the lower glass (red curve) slide.}
\label{fig1}
\end{figure}

\paragraph{Velocity field at the boundary}
Fig.\ \ref{fig2}a,b) give a schematic view of the velocity profile which saturates at distances beyond  $\lambda$ in the effective slip velocity $v_{s}$ (eq. \ref{20}). As the temperature gradient in our setup is not constant along the solid boundary one expects that the slip velocity along the upper plate decreases with the inverse square of the radius,
\begin{equation}
	v_{s}=-\chi\frac{\Delta T_{\rm{Au}}}{T}\frac{a}{r^{2}}. \label{32}
\end{equation}
For $\chi>0$ the surface flow is oriented towards the origin. A corresponding but considerably weaker surface flow ($v'_{s}$) is also present at the other glass cover slide. These surface flows induce a radially symmetric volume flow ${\bf w}(r,z)$, which is traced by the gold nano particles.

\begin{figure}[ptb]
\begin{center}
\includegraphics[width=0.7 \columnwidth]{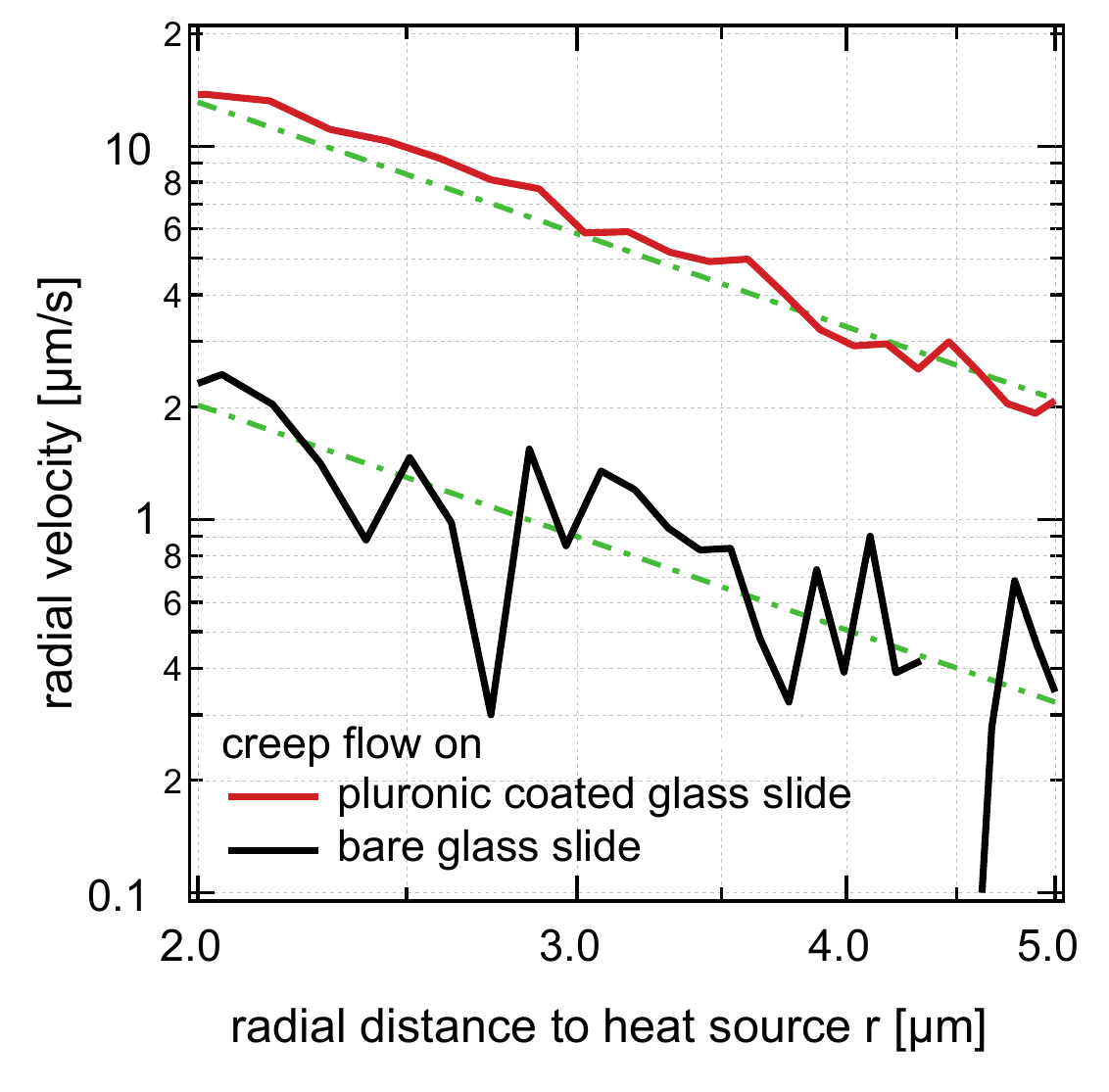}
\caption{Temperature profile and slip velocity along the upper boundary. The temperature agrees very well the power law $\propto r^{-2}$. The slip velocity on both glass and Pluronic-coated glass follows this law beyond 2 microns from the heated spot.}
\label{fig3}
\end{center}
\end{figure}

In Fig.\ \ref{fig3} we plot the experimentally obtained velocity of the tracers close to the upper boundary of the liquid film as a function of the radial distance from the heated gold nanoparticle. The velocity profile has been measured for a bare glass plate (black curve) and one coated with Pluronic F-127 (red curve). At distances larger than 2 microns from the heat source, both agree well with the power law $\propto r^{-2}$ (indicated by the green dashed lines). For both systems, the flow is towards higher temperatures. The measured velocities differ by a factor of 7.5 where the larger one is found for the Pluronic F-127 coated surface. This suggests that the mobility parameter $\chi$ is considerably stronger for the non-ionic block-copolymer as compared to the charged bare glass surface. A value for the thermo-osmotic surface velocity $v_{s}$ can be extracted when considering that the tracers velocity represents an average $\langle {\bf w}(r,z)\rangle _{\Delta r}$ over twice the diffusion length ($\Delta r=620\, nm$) during the exposure time (see SI\cite{SI} for details of the averaging). According to this we find a maximum velocity of $v_{s}^{\rm glass}=40\ \mu m/s$ and $v_{s}^{\rm  Pluronic}=300\ \mu m/s$

\paragraph{Thermo-osmotic coefficient $\chi$.} The surface velocities and the temperature increments $\Delta T_{\rm{Au}}$ yield the thermo-osmotic coefficients. Table 1 compares the coefficient $\chi$ obtained for the Pluronic coated and bare glass slides, with previously reported thermo-osmotic data and the reduced thermophoretic mobility $TD_T$ of colloidal suspensions. For large particles the coefficients are related through $TD_T={2\over3}\chi$ \cite{And89}, whereas for small particles and polymers one rather has $TD_T\simeq\chi$.  

Pluronic F-127 is a non-ionic triblock copolymer consisting of a central hydrophobic block of polypropylene glycol (PPG, 56 repeat units, bright yellow in Fig. 1 b) and two hydrophilic blocks of poly-ethylene glycol (PEG, 101 units, dark red). Both molecules show complex thermodynamic behavior in water; PEG  is soluble under the present conditions, whereas PPG is not \cite{Kje81}. Thus the PPG block is assumed to stick to the glass slides, whereas the more hydrophilic PEG parts form a molecular brush, as illustrated in Fig. 1 b). As a consequence, there is no well defined plane-of-shear. 

\begin{table}[t]
\caption{Thermo-osmotic parameter $\chi$. Our results are compared with data of previous work for both $\chi$ and the reduced thermophoretic mobility $TD_T$, and with theoretical values as discussed in the main text. $D_T$ is obtained from Soret data for poly-ethylene glycol \cite{Cha03,Kit04} and Ludox particles \cite{Nin08}. The  scatter of previous data is due to differences between porous glass specimens \cite{Der87} and the dependence of $D_T$ on pH and salinity \cite{Nin08}.  We also mention that $D_T$ of PEG depends on both molecular weight \cite{Cha03} and temperature \cite{Kit04}.}
\begin{tabular}
[c]{|l|c|c|c|c|}\hline
 $\chi$ ($10^{-10}$m$^{2}$/s)    & this work &  theory & prev. work   &  $TD_T$  \\\hline
Pluronic F-127 & 13                      &           $\sim 14$  &                              & 15 \cite{Cha03,Kit04} \\\hline
Glass    &  1.8                    &       $\sim 1$      & $-(0.2...1.5)$ \cite{Der87} & $-(1...9)$ \cite{Nin08} \\\hline
\end{tabular}
\end{table}
 
Since the water-polymer interface consists essentially of hydrophilic blocks, one expects the parameter $\chi$ to be mainly determined by the PEG properties. The excess enthalpy of PEG in water results from the balance of the hydrogen bridging of the oxygen ($h<0$) and the opposite hydrophobic effect of the poly-ethylene. The measured enthalpy of mixing at low PEG content is $\Delta H=-0.66\times10^{-20}$ J per monomer \cite{Mal57}. Taking the polymer as a rod of radius $b$ and unit length $d$, and assuming that the enthalpy density $h$ is constant within the interaction length $\lambda$ and zero beyond, we obtain $h=\Delta H /2\pi b d\lambda$ and a thermo-osmotic coefficient
   \begin{equation} 
  \label{eq42}
   \chi=-\frac{\Delta H}{\eta}\frac{\lambda}{4\pi b d} \sim 14\times 10^{-10} {\rm m^2}/{\rm s},
   \end{equation}
with $b=\lambda$ and the numerical value $d=3.5$\AA. This estimate agrees well with the present measurement and previous thermophoresis data for PEG. 

The situation is less clear concerning thermo-osmosis on a bare glass surface. Our experiment shows a slip velocity towards the heated spot, implying a negative enthalpy $h$, and the analysis of the experimental data gives $\chi=1.8 \times 10^{-10}\,\rm{m^2/s}$. Previous works, however, reported the opposite effect: In their thermo-osmosis experiments on porous glass, Derjaguin et al. found $\chi<0$ \cite{Der87}, and the Soret data of Rusconi et al. on Ludox particles provided a positive thermophoretic mobility $D_T$ \cite{Nin08}.  

The excess enthalpy $h$ in the vicinity of a glass surface is a complex quantity. Besides electrostatic and van der Waals forces, structuration of water seems to play a major role \cite{Der80}. The latter results in an increase of enthalpy ($h>0$), whereas the electric-double layer contribution is negative. This would suggest that the positive coefficient $\chi$ of the present experiment results from surface charges, and that previous findings are dominated by a strong structuration contribution ($\chi<0$). This is supported by the temperature dependence of Derjaguin's data \cite{Der87}: The negative thermo-osmotic coefficient $\chi$ has a large positive derivative $d\chi/dT>0$; this is expected for the structuration of water which is strong at low temperatures yet weakens at higher $T$. On the other hand, the dependence of $D_T$ on salinity and pH reported in \cite{Nin08}, indicates a significant electrostatic contribution, in addition to the dominant structuration effect. 

The above discussion suggests that the present data ($\chi>0$) are related to the electrostatic contribution. Glass in contact with water in general carries surface charges, which give rise to an electric double layer with Debye screening length $\lambda$, as sketched in Fig.\ \ref{fig2}. The electric-double layer enthalpy is negative and thus results in a flow towards the heat source ($v_{s}<0$).\ In the Debye-H\"uckel approximation one has $h=-\frac{1}{2}\varepsilon(\zeta/\lambda)^{2} e^{-2z/\lambda}$, where $\varepsilon$ is the permittivity and $\zeta$ the surface potential, resulting in
   \begin{equation} 
   \chi=\frac{\varepsilon\zeta^{2}}{8\eta}. 
   \end{equation}
With $\zeta\sim 30\,\rm{mV}$, one finds $\chi\sim 10^{-10}\,\rm{m^2/s}$, which agrees with both the sign and the order of magnitude of the measured value. 

\paragraph{Bulk velocity field}

\begin{figure}[ptb]
\includegraphics[width=0.8\textwidth]{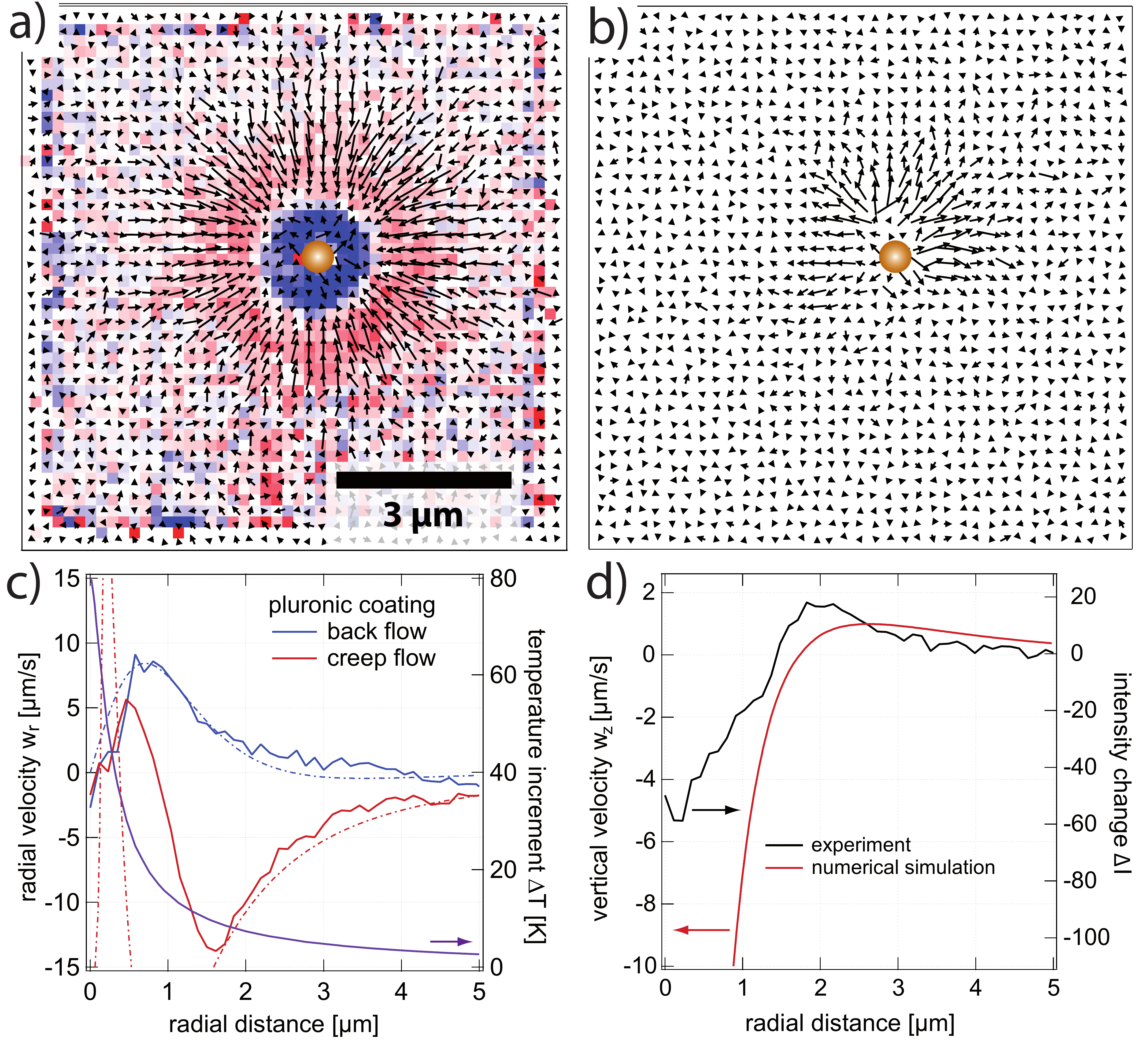}
\caption{
 \textbf{a)} Two dimensional velocity map of the tracer motion at the upper surface. The color indicates the vertical velocity with blue being a flow in the negative z-direction away from the heat source and red in the positive z-direction towards the upper glass slide where the heat source is located at. \textbf{b)} Shows the two dimensional velocity map of the backflow within the sample. \textbf{c)}  Displays the average radial velocity at the upper surface (red curve) and the backflow (blue curve) as a function of the radial distance to the heat source. Negative velocities represent thereby a flow towards the heat source. The dashed line displays the vertically averaged radial velocity obtained from numerical calculations. In addition the temperature increment in the surrounding of the heat source is displayed (purple curve). \textbf{d)} The average change of the tracer intensity and the corresponding vertical velocity in comparison to the numerical calculation.}
\label{fig4}
\end{figure}

In order to fully characterize the thermo-osmotic flow field, we measured both radial and vertical components $\rm{w}_{\rm{r}}$ and $\rm{w}_{\rm{z}}$ of the bulk velocity field ${\bf w}(r,z)$ and compare it to numerical results using finite element simulations (COMSOL)\cite{SI}. The vertical separation is achieved by collecting signals of different intensity, which correspond to tracer particles at different $z$. Figure \ref{fig4} a) depicts the flow field detected for the brightest tracer particles at the interface, which is directed towards the heat source. At intermediate intensities the flow is directed away from the heat source (Fig. \ref{fig4} b). This reversal of the flow direction is obvious from the mass conservation; the inward slip velocity at the boundaries requires an outward flow in the center of the liquid film. The corresponding radial dependencies are plotted in Fig. \ref{fig4} c) and agree well with the $1/r^2$ dependence expected from eq. \ref{32} at distances beyond $2\,\rm{\mu m}$. At shorter distances, the velocity profile reveals a vertical component $w_{z}$. This vertical component, averaged over the complete film thickness, is detected by counting the number of particles which decrease or increase their scattering intensity within neighboring frames. A decrease in intensity corresponds to a downward motion of the particles, away from the heat source, while an increase intensity reveals an upward motion. Figure \ref{fig4} a) depicts in the color underlay that close to heat source particles on average move vertically away from the heat source (blue). At about $r=2\, \mu m$ distance from the heat source a motion in the opposite direction is observed, while even farther away no net transport in the vertical direction is found. The corresponding intensity changes are plotted in Figure \ref{fig4} d). This detected vertical flow can only be due to the thermo-osmosis as convection is in the used sample geometry absent. Further, the radiation pressure acting in the region of the heating laser is pointing against the detected vertical flow. The complete calculated flow field is depicted in  Fig.\ \ref{fig2} c). 

\paragraph{Summary}
We have measured the velocity field caused by thermo-osmosis due to a heated gold nanoparticle fixed at the surface of a glass slide. The inferred interfacial flow velocities reach values of up to $300\, \mu m/s$ and set up a parabolic flow field in the water film at large distances from the heat source. The slip velocities are found much stronger at interfaces covered with non-ionic block-copolymers as compared to bare charged glass interfaces. 
The good agreement of the numbers gathered in \mbox{Table 1} leads us to the conclusion that the slip velocity on a Pluronic-coated surface is driven by thermo-osmosis on the PEG-water interface. More generally, they confirm the role of the interaction enthalpy for the thermal separation of molecular mixtures \cite{Wue14} and suggest that simple models such as eq. \ref{eq42} provide a good description for Soret data of polymers. Regarding the thermo-osmosis on a glass surface, our findings differ from previous results. The discrepancy probably results from the competition of electric-double layer forces and structuration of water; their contributions to the excess enthalpy are of opposite sign and thus may result in a positive or negative slip velocity.   
Our results further suggest that thermo-osmotic flows along solid/liquid interfaces may contribute considerably to thermophoretic measurements in thin film geometries and may be harnessed for microfluidic applications.

The authors acknowledge financial support by Sonderforschungsbereich TRR 102, the DFG priority program 1726 "Microswimmers", the S\"achsische Forschergruppe FOR 877, the DFG the joint DFG/ANR project "Thermoelectric effects at the nanoscale" and Leibniz Program at the Universit\"at Leipzig.


\begin{thebibliography}{99}                                                                                               %

\bibitem{Log12} B.E. Logan, M. Elimelech, Nature \textbf{488}, 313 (2012).
\bibitem{Pic13} C.B. Picallo, S. Gravelle, L. Joly, E. Charlaix,  L. Bocquet, Phys. Rev. Lett. \textbf{\ 111}, 244501 (2013).
\bibitem {Der41}B.V. Derjaguin,. G.P. Sidorenkov, Doklady Akad. Nauk. SSSR \textbf{32}, 622 (1941).
\bibitem {Jia10}H.-R. Jiang, N. Yoshinaga, M. Sano, Phys. Rev. Lett. \textbf{\ 105}, 268302 (2010).
\bibitem {But12}I. Buttinoni, G. Volpe, F. K\"{u}mmel, G. Volpe, C. Bechinger, J. Phys.: Condens. Matter \textbf{24}, 284129 (2012).
\bibitem{Qia13} B. Qian, D. Montiel, A. Bregulla, F. Cichos, H. Yang, Chem. Sci. \textbf{4}, 1420 (2013).
\bibitem{Wei08} F. Weinert, D. Braun, Phys. Rev. Lett. Phys. Rev. Lett. \textbf{101}, 168301 (2008).
\bibitem{Leo09} R. Di Leonardo, F. Ianni, G. Ruocco, Langmuir \textbf{25}, 4247 (2009).
\bibitem {Bre14}A. Bregulla, H. Yang, F. Cichos, ACS Nano \textbf{8}, 6542 (2014).
\bibitem {Duh06}S.\ Duhr and D.\ Braun, Phys. Rev. Lett. \textbf{97}, 038103 (2006).
\bibitem {Mae11}Y.T. Maeda, A. Buguin, A. Libchaber, Phys. Rev.Lett. \textbf{107}, 038301 (2011).
\bibitem {Ped14}J.N. Pedersen, C.J. L\"oscher, R. Marie, L.H. Thamdrup, A. Kristensen, H. Flyvbjerg, Phys. Rev. Lett. \textbf{113}, 268301 (2014).
\bibitem {Bra13}M.\ Braun, F.\ Cichos, ACS Nano \textbf{7}, 11200 (2013).
\bibitem {Bra15}M.\ Braun,A.\ Bregulla, K.\ G\"unther, M.\ Mertig, F.\ Cichos, Nano Lett. \textbf{15},  5499 (2015).
\bibitem {Yu15}L.-H.  Yu,  Y.-F.  Chen, Anal. Chem. \textbf{87}, 2845 (2015).
\bibitem {Wue10}A.\ W\"{u}rger, Rep. Prog.\ Phys. \textbf{73}, 126601(2010).
\bibitem {Wue15}A.\ W\"{u}rger, Phys. Rev. Lett. \textbf{115}, 188304 (2015).
\bibitem{Esl14}K.A. Eslahian, A. Majee, M. Maskos, A. W\"urger, Soft Matter {\bf 10}, 1931 (2014).
\bibitem{Vig10} D. Vigolo, S.\ Buzzaccaro and R.\ Piazza, Langmuir \textbf{\ 26}, 7792 (2010).
\bibitem{Bro14} A. Brown, W. Poon, Soft Matter \textbf{10}, 4016 (2014).
\bibitem{Rei14}M. Reichl, M. Herzog, A. G\"otz, D. Braun, Phys. Rev. Lett. {\bf 112}, 198101 (2014).

\bibitem{Chen15} J. Chen, Z. Kang, S.K. Kong, H.-P. Ho, Opt. Lett. \textbf{40}, 3926 (2015).
\bibitem{Cuche13} A. Cuche, A. Canaguier-Durand, E. Devaux, J.A. Hutchison, C. Genet, T.W. Ebbesen, Nano Lett. \textbf{13}, 4230 (2013).

\bibitem {Der87}B.V.\ Derjaguin, N.V.\ Churaev, V.M.\ Muller, Surfaceforces Plenum New York (1987).
\bibitem {Vil06} J.P.G. Villaluenga, B. Seoane, V.M. Barrag\'an, C. Ruiz-Bauz\'a, J.  Membrane Sci. \textbf{274}, 116 (2006).
\bibitem {Kim09} S. Kim, M.M. Mench, J. Membrane Sci. \textbf{328}, 113 (2009).
\bibitem {Gro62}S.R. de Groot, P.\ Mazur, Non-equlibrium Thermodynamics, North Holland Publishing, Amsterdam (1962).
\bibitem {SI} See Supplemental Material
\bibitem {And89} J.L. Anderson, Ann. Rev. Fluid Mech. \textbf{21}, 61 (1989).
\bibitem {Kje81} R. Kjellander, E. Florin, J. Chem. Soc., Faraday Trans. 1  \textbf{24}, 2053 (1981).
\bibitem {Mal57} G.N. Malcolm, J.S. Rowlinson, Trans. Faraday Soc. \textbf{53}, 921 (1957).
\bibitem {Cha03}J. Chan, J.\ Popov, S.\ Kolisnek-Kehl, D. Leaist, J. Solut.Chem. \textbf{32}, 197 (2003).
\bibitem {Kit04}R. Kita, S. Wiegand, Jutta Luettmer-Strathmann,J.\ Chem.\ Phys. \textbf{121}, 3874 (2004).

\bibitem {Nin08} R. Rusconi, L. Isa, and R. Piazza, J. Opt. Soc. Am. B \textbf{21}, 605 (2004).
\bibitem {Der80}B.V.\ Derjaguin, Pure Appl. Chem. {\bf 52}, 1163 (1987).

\bibitem {Wue14}A.\ W\"{u}rger, J. Phys. Condens. Matt.  \textbf{26}, 035105 (2014).

\end{thebibliography}
\end{document}